\documentclass{elsartL}
\pdfoutput=1
\usepackage{graphicx}
\usepackage{amssymb}
\usepackage{amsmath}
\usepackage{epstopdf}
\begin{document}

\begin{frontmatter}
\title{Processing of acoustical signals via a wavelet-based analysis}
\author[IMS]{E.~Matsinos{$^*$}}
\address[IMS]{Institute of Mechatronic Systems, School of Engineering, Zurich University of Applied Sciences (ZHAW), Technikumstrasse 5, CH-8401 Winterthur, Switzerland}

\begin{abstract}
In the present paper, details are given on the implementation of a wavelet-based analysis tailored to the processing of acoustical signals. The family of the suitable wavelets (`Reimann wavelets') are obtained in the time domain 
from a Fourier transform, extracted in Ref.~\cite{r1} after invoking theoretical principles and time-frequency localisation constraints. A scheme is set forth to determine the optimal values of the parameters of this type of 
wavelet on the basis of the goodness of the reproduction of a $30$-s audio file containing harmonic signals corresponding to six successive $A$ notes of the chromatic musical scale, from $A_2$ to $A_7$. The quality of the 
reproduction over about six and a half octaves is investigated. Finally, details are given on the incorporation of the re-assignment method in the analysis framework, as the means a) to determine the important contributions of 
the wavelet transforms and b) to suppress noise present in the signal.
\end{abstract}
{$^*$}{E-mail: evangelos.matsinos[AT]zhaw.ch}
\end{frontmatter}

\section{\label{sec:Introduction}Introduction}

The analysis of acoustical signals and visual images in terms of wavelet transformations is a `multi-resolution analysis'; as such, it purports at providing a solution to time-frequency localisation considerations arising in the 
standard `short-time Fourier transformation' (STFT), namely to the deterioration of the frequency resolution in compactly supported (i.e., associated with narrow time windows) STFT analyses. A wavelet-based analysis (WBA) probes 
the input signal at various scale levels, analogous to the frequency levels in the standard Fourier-based analysis (FBA), matching the resolution requirements associated with each such level.

The involvement of the wavelet transformations in understanding the cochlear response to external stimuli was proposed in the early 1990s \cite{dob,yws}. The relevance of such an analysis in this subject appertains to the near 
fulfillment of the local scaling symmetry, under which the cochlear filter is a function of one variable, i.e., of the ratio of the input angular frequency $\omega$ to the tonotopic axis, which associates positions along the 
basilar membrane of the cochlea with frequencies \cite{zwg}. Known as Zweig's postulate, this property had been established in von B{\'e}k{\'e}sy's early experiments \cite{bek}, which demonstrated that the graphs of the cochlear 
filter at two (different) frequencies are translated against one another by a constant term involving the logarithm of the ratio of the two frequencies. Given the potential for commercial applications, it is hardly surprising that 
aspects of this mathematical subject have been patented \cite{bt}.

The Fourier transform of the family of wavelets, which have been put forth as suitable candidates for the processing of acoustical signals, was obtained in Ref.~\cite{r1} on the basis of theoretical principles and of the 
time-frequency localisation constraints. Herein, each such wavelet is called a `Reimann wavelet'. The goal in this paper is to show that the wavelet transformations, emerging from this family, establish a promising theoretical 
basis for the processing of acoustical signals. If successful in this task, the possibility arises of using these wavelet forms in commercial applications, even to the extent of replacing the Fourier transformation upon which 
the processing of acoustical signals predominantly relies.

This paper is organised as follows.
\begin{itemize}
\item In Section \ref{sec:Method}, the relevant definitions are given and the mathematical background, required in the processing of acoustical signals, is developed. Subsection \ref{sec:WaveletTransform} deals with general 
features of the continuous wavelet transformations, whereas Subsection \ref{sec:WaveletTimeDomain} focusses on the specific family of wavelets introduced in Ref.~\cite{r1}. Subsection \ref{sec:SERe} explores specific properties 
of the transforms, obtained with the family of the Reimann wavelets, in particular regarding the `structure equations', i.e., relations which these transforms obey; the last part of the subsection investigates the 
incorporation of the re-assignment method in the analysis, aiming at providing solutions to two problems: to the determination of the important contributions of the wavelet transforms and to the development of an efficient 
noise-suppression scheme.
\item In Section \ref{sec:Results}, the results of the application of the theoretical background of Section \ref{sec:Method} are discussed. The first part of the section pertains to the settings entering the analysis of the 
signals, whereas Subsection \ref{sec:Parameters} outlines the strategy leading to the determination of optimal values for the parameters of the Reimann wavelets. In Subsection \ref{sec:Harmonic}, investigated is the goodness of 
the reproduction of a few harmonic signals spanning about six and a half octaves, from $80$ to $7\,040$ Hz. The effects of the noise and options for the noise suppression are touched upon in Subsection \ref{sec:Noise}.
\item A summary of the findings, obtained in the present study, is provided in Section \ref{sec:Conclusions}, and a few directions for future research are delineated.
\end{itemize}

\section{\label{sec:Method}Method}

\subsection{\label{sec:WaveletTransform}Definition of the continuous wavelet transformation}

The wavelet transform of a continuous, square-integrable function $f \in \mathbb{L}^2 (\mathbb{R})$ is defined as
\begin{equation} \label{eq:EQ0010}
WT_f^\psi (s, \tau) = \frac{1}{\sqrt{\lvert s \rvert}} \int_{-\infty}^{\infty} f(t) \psi^* \left( \frac{t-\tau}{\lvert s \rvert} \right) dt \, \, \, ,
\end{equation}
where $t$ denotes the time, and the real variables $s$ and $\tau$ are respectively known as scaling factor and time-shift factor. The scaling factor $s$ is dimensionless and is assumed positive herein; the time-shift factor $\tau$ 
has the dimension of time and may be either positive or negative. The asterisk in Eq.~(\ref{eq:EQ0010}) indicates complex conjugation. After restricting $s$ to positive values, one obtains
\begin{equation} \label{eq:EQ0011}
WT_f^\psi (s, \tau) = \frac{1}{\sqrt{s}} \int_{-\infty}^{\infty} f(t) \psi^* \left( \frac{t-\tau}{s} \right) dt \, \, \, .
\end{equation}
The function $\psi (t)$, which is assumed continuous both in the time and the frequency domains, is the `mother wavelet'. By variation of $s$ and $\tau$, different forms (i.e., the daughter wavelets) are generated. One 
may think of the daughter wavelets as functions of $t$, with parameters $s$ and $\tau$, according to the expression
\begin{equation*}
\psi_{s \tau}(t) = \frac{1}{\sqrt{s}} \psi \left( \frac{t-\tau}{s} \right) \, \, \, .
\end{equation*}

The inverse transformation is given by the expression
\begin{equation} \label{eq:EQ0013}
f(t) = \frac{2}{c_{\psi}^2} \int_{0}^{\infty} \frac{ds}{s^{5/2}} \int_{-\infty}^{\infty} WT_f^\psi (s, \tau) \psi \left( \frac{t-\tau}{s} \right) d\tau \, \, \, ,
\end{equation}
where
\begin{equation} \label{eq:EQ0015}
c_{\psi}^2 = 2 \pi \int_{-\infty}^{\infty} \frac{\lvert \hat{\psi} (\omega) \rvert^2}{\lvert \omega \rvert} d\omega \, \, \, .
\end{equation}
In the last equation, $\hat{\psi} (\omega)$ stands for the Fourier transform of the mother wavelet $\psi (t)$. Evidently, the inverse wavelet transformation is possible only if the admissibility constant $c_{\psi}^2$ fulfills the 
condition
\begin{equation} \label{eq:EQ0020}
0 < c_{\psi}^2 < \infty \, \, \, .
\end{equation}

If a process involves both the direct and the inverse wavelet transformations, the absolute normalisation of the wavelet drops out. However, in order to compare wavelet forms and transforms obtained with different parameter sets, 
the wavelets need to be properly normalised. The normalisation factors are obtained via the imposition of the condition
\begin{equation*}
\int_{-\infty}^{\infty}\lvert \psi(t) \rvert^2 dt \equiv \int_{-\infty}^{\infty} \lvert \hat{\psi} (\omega) \rvert^2 d\omega = 1 \, \, \, . 
\end{equation*}

\subsection{\label{sec:WaveletTimeDomain}The Reimann wavelets}

The general expression for the Fourier transform of the wavelet family, which has been proposed as suitable for the modelling of the signal processing in the cochlea, was derived in Ref.~\cite{r1} on the basis of local-invariance 
principles, as well as of the limitations in the localisation of the signal in the time and the frequency representations \cite{r2}; the analogue in Quantum Physics of the localisation constraints is the Heisenberg uncertainty 
principle. The form of the Reimann wavelets in the frequency domain is
\begin{equation} \label{eq:EQ0030}
h(y) = k \, \exp (i \phi(y)) \, \exp (-\kappa \lvert y \rvert^c / c) \, \lvert y \rvert^{\kappa \nu - 1/2} \, \, \, ,
\end{equation}
where the argument $y$ is proportional to the input angular frequency $\omega$; $\kappa$, $\nu$, and $c$ are positive parameters. The absolute normalisation $k \in \mathbb{R}^+$ is obtained from the parameters $\kappa$, $\nu$, 
and $c$ via the expression
\begin{equation} \label{eq:EQ0035}
k = \left( \frac{2 \kappa}{c} \right)^{\kappa \nu/c} \sqrt{\frac{c \pi}{\Gamma(\frac{2 \kappa \nu}{c})}} \, \, \, ,
\end{equation}
where $\Gamma (n)$ is the complete gamma function.

The phase $\phi(y)$ is given by
\begin{equation} \label{eq:EQ0040}
\phi(y) = \epsilon \, {\rm sgn}(y) + \alpha \, {\rm sgn}(y) \ln\lvert y \rvert - \beta y \, \, \, .
\end{equation}
Three parameters ($\epsilon$, $\alpha$, and $\beta$) are introduced via Eq.~(\ref{eq:EQ0040}); $\alpha$ and $\beta$ are assumed positive~\footnote{In Ref.~\cite{r1}, the parameter $\alpha$ is assumed negative; this difference in 
the sign convention ought to be remembered when comparing expressions (and values) taken from Ref.~\cite{r1} with those of the present paper.}.

The modes of the envelope of $h(y)$ are given by the expression
\begin{equation} \label{eq:EQ0041}
\lvert y_e \rvert = \left( \nu - \frac{1}{2 \kappa} \right)^{1/c} \, \, \, .
\end{equation}
One interesting relation is
\begin{equation*}
\frac{\int_{-\infty}^{\infty} \lvert y \rvert^c \lvert h(y) \rvert^2 dy}{\int_{-\infty}^{\infty} \lvert h(y) \rvert^2 dy} = \nu \, \, \, ;
\end{equation*}
therefore, the parameter $\nu$ is equal to the expectation value of $\lvert y \rvert^c$. It is due to this relation that the parameter $\nu$ is rather considered to be a normalisation constant in Ref.~\cite{r1}, and its value is 
set to $1$ therein. Later on, $\nu$ will also be set to $1$ in this paper.

The relation between the variable $y$ and the input angular frequency $\omega$ involves the tonotopic axis (or frequency-position map) $\xi(x)$, which has the dimension of angular frequency;
\begin{equation} \label{eq:EQ0045}
y = \frac{\omega}{\xi(x)} \, \, \, .
\end{equation}
It is known that high frequencies are processed at the basis of the human cochlea, whereas low ones are processed in the apical region. The quantity $x$ in the previous equation may be defined as the fraction of the length of the 
basilar membrane (which in humans is around $35$ mm), so that the basis of the cochlea corresponds to $x = 0$ and the end of the cochlear apex to $x = 1$. Therefore, the function $\xi(x)$ fulfills the conditions: 
$\xi(0) = \omega_{\rm max}$ and $\xi(1) = \omega_{\rm min}$, where $\omega_{\rm max}$ and $\omega_{\rm min}$ represent the extreme values of the angular frequency of acoustical signals which may be processed in the cochlea. 
Although it is generally claimed that $\omega_{\rm max}$ and $\omega_{\rm min}$ may be respectively set to $2 \pi 20\,000$ rad/s and $2 \pi 20$ rad/s for humans, a slightly larger $\omega_{\rm min}$ threshold is adopted herein, 
namely $2 \pi 60$ rad/s. An exponential relation may be introduced
\begin{equation*}
\xi (x) = \omega_{\rm max} e^{-\gamma x} \, \, \, ,
\end{equation*}
where the parameter $\gamma$ is obtained from the aforementioned values of $\omega_{\rm max}$ and $\omega_{\rm min}$: $\gamma \approx 5.81$. Invoking the scaling factor $s$, one may put Eq.~(\ref{eq:EQ0045}) into the convenient 
form
\begin{equation*}
y = \frac{s \omega}{\omega_0} \, \, \, ,
\end{equation*}
where $\omega_0$ is a characteristic angular frequency, with respect to which the scaling factor $s$ is defined. (Note that the scaling factor $s$ appears as $a$ in Ref.~\cite{r1} and that it is not dimensionless therein.) In this 
paper, $\omega_0$ is taken to be the angular frequency of the note $A_5$, corresponding to $880$ Hz, i.e., close to the geometrical mean of $\omega_{\rm max}$ and $\omega_{\rm min}$ for humans; thus, $\omega_0 = 2 \pi 880$ rad/s. 
It is useful to write Eq.~(\ref{eq:EQ0030}) in the form:
\begin{equation} \label{eq:EQ0046}
h(s \omega) = \frac{k}{\omega^{\kappa \nu}_0} \, \exp \left( i \phi(\frac{s \omega}{\omega_0}) \right) \, \exp \left( -\frac{\kappa}{c} \left| \frac{s \omega}{\omega_0} \right|^c \right) \, \lvert s \omega \rvert^{\kappa \nu - 1/2} \, \, \, ,
\end{equation}
and use this definition of the function $h(s \omega)$ henceforth. This expression may be easily obtained after considering the mapping $\lvert h(y) \rvert^2 dy \to \lvert h(s \omega) \rvert^2 d(s \omega)$.

The relation between the Fourier transform $\hat{\psi}_{s \tau}(\omega)$ and the function $h(s \omega)$ of Eq.~(\ref{eq:EQ0046}) may be found in Ref.~\cite{r1}:
\begin{equation} \label{eq:EQ0047}
\hat{\psi}_{s \tau} (\omega) = \frac{\sqrt{s}}{\sqrt{2 \pi}} h^*(s \omega) e^{- i \omega \tau} \, \, \, .
\end{equation}
The application of the inverse Fourier transformation generates a real wavelet
\begin{align} \label{eq:EQ0050}
\psi_{s \tau} (t) & = \frac{1}{\sqrt{2 \pi}} \int_{-\infty}^{\infty} \hat{\psi}_{s \tau} (\omega) e^{i \omega t} d\omega\nonumber\\
&= \frac{k s^{\kappa \nu}}{\pi \omega^{\kappa \nu}_0} \int_{0}^{\infty} \cos \left( \omega t - \phi(\frac{s \omega}{\omega_0}) - \omega \tau \right) \, \exp \left( -\frac{\kappa s^c}{c\omega^c_0} \omega^c \right) \, \omega^{\kappa \nu - 1/2} d\omega \, \, \, .
\end{align}
To obtain this expression, use has been made of the antisymmetry of $\phi(s \omega / \omega_0)$, i.e., of the relation $\phi(-y) = -\phi(y)$. To set the wavelet (and, thus, be able to perform the direct and inverse transformations), 
one must evaluate the integral of an oscillating function. This issue will be addressed shortly. The mother wavelet in the time domain may be obtained from Eq.~(\ref{eq:EQ0050}) after setting $s=1$ and $\tau=0$ s.

Let us now have a closer look at the properties of $\phi(y)$ of Eq.~(\ref{eq:EQ0040}). This function is continuous everywhere, save for $y=0$ (where it is not defined). Given that the integration in Eq.~(\ref{eq:EQ0050}) is 
performed from $0$ to $\infty$, one is basically left with three options in the implementation. a) Evaluate the integral of Eq.~(\ref{eq:EQ0050}) after implementing a method suitable for integration in case of open intervals, 
e.g., a variant of Romberg's integration method \cite{ptvf}. b) Make use of Eq.~(\ref{eq:EQ0050}), but replace the logarithmic function of Eq.~(\ref{eq:EQ0040}) by a constant value below a user-defined threshold; in practice, 
this may be easily achieved via the redefinition of the logarithmic function. The drawback of these two options is that they generate a phase which is not continuous at $y=0$. c) Replace the logarithmic function (below some 
threshold) with a function passing through the origin ($0$,$0$). The choice of the form is arbitrary, but one may opt for the simplest solution, i.e., for a linear function in the vicinity of $y=0$. Method (c), our preference 
herein, generates a phase which is continuous at $y=0$.

\begin{figure}
\begin{center}
\includegraphics[width=15.5cm]{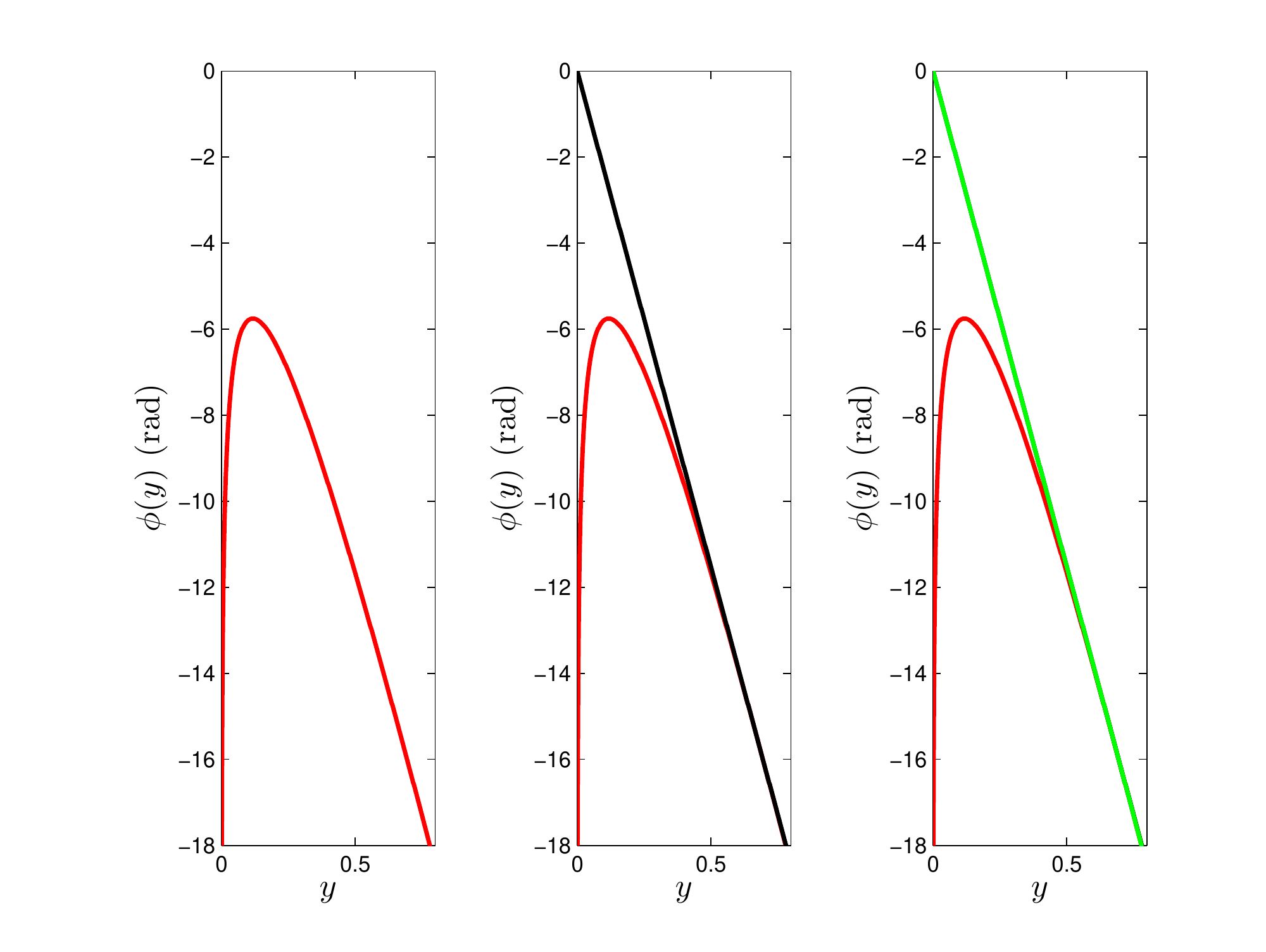}
\caption{\label{fig:phase}Plot of the function $\phi(y)$ of Eq.~(\ref{eq:EQ0080}) for the optimal values of the parameters $\alpha$, $\beta$, and $\phi_m$ from Table \ref{tab:OptParVal}. The black straight line in the middle 
plot passes through the origin ($0$,$0$) and is tangent to the curve at $y=y_t$. The curve in green, representing the simplest solution without a `kink', is identical to the tangent (black straight line) for $0 \leq y \leq y_t$ 
and to the original form (curve in red) for $y>y_t$. (The black and green curves almost coincide in the $y$ domain of the plot.)}
\vspace{0.35cm}
\end{center}
\end{figure}

It must be remembered that $\phi(y)$, as given in Eq.~(\ref{eq:EQ0040}), serves as an approximation in modelling a physiological process; the available data do not support the $\phi(y)$ behaviour of Eq.~(\ref{eq:EQ0040}) in the 
vicinity of $y=0$. As $\phi(y)$ is an antisymmetric function, it suffices to examine its properties for $y>0$. In this case, Eq.~(\ref{eq:EQ0040}) is written as
\begin{equation} \label{eq:EQ0060}
\phi(y) = \epsilon + \alpha \ln y - \beta y \, \, \, .
\end{equation}
The maximum $\phi_m$ of this function occurs at $y_m=\alpha / \beta$ and may be put in the form
\begin{equation} \label{eq:EQ0070}
\phi_m = \epsilon + \alpha ( \ln y_m - 1 ) \, \, \, .
\end{equation}
By using Eqs.~(\ref{eq:EQ0060},\ref{eq:EQ0070}), one eliminates the parameter $\epsilon$.
\begin{equation} \label{eq:EQ0080}
\phi(y) = \phi_m + \alpha \left( \ln \frac{y}{y_m} + 1 - \frac{y}{y_m} \right)
\end{equation}
A typical plot of $\phi(y)$ is shown in Fig.~\ref{fig:phase}. The $y$ value of the point at which the straight line, drawn from the origin ($0$,$0$), is tangent to this curve is given by: $y_t=y_m e^{-\phi_m / \alpha}$; the slope 
of this tangent line is equal to $\beta (e^{\phi_m / \alpha} - 1)$. The simplest solution for $\phi(y)$ without a `kink' (i.e., with continuous first derivative) is also shown in the figure (curve in green); this form is continuous 
at $y=0$, being identified with the tangent to the original curve of Eq.~(\ref{eq:EQ0040}) below $y_t$ and with the original curve above $y_t$. This is our recommendation for the function $\phi(y)$.

As (for large values of $\lvert t \rvert$) the integrand of Eq.~(\ref{eq:EQ0050}) is rapidly oscillating, performing the numerical integration reliably is not trivial. The direct application of Romberg's algorithm (which is not 
expected to be a suitable method for such an integration) yielded results which showed persistent `noise' at large $\lvert t \rvert$. A number of transformations were subsequently attempted, but failed to yield results better 
than those obtained with the direct integration.

The problem with rapidly oscillating integrands is that the standard integration algorithms fail to follow the rapidity of the oscillations. (In this case, aliasing might also become relevant.) The present study circumvents this 
problem by performing the integration within successive roots of the function which is responsible for the oscillatory behaviour of the integrand, namely of the cosine function of Eq.~(\ref{eq:EQ0050}); these elementary integrals 
are evaluated reliably. The roots of the cosine function may be evaluated by setting
\begin{equation*}
y \omega_0 t - \phi(y) = \frac{\pi}{2} + n \pi \, \, \, ,
\end{equation*}
where $n \in \mathbb{Z}$. In fact, the accurate determination of the roots is inessential; the integration intervals are treated in a sequential manner, which implies that any root appears twice in the series of the elementary 
contributions, once as an upper integration limit, once as a lower integration limit. Finally, by summing up these elementary contributions, one obtains an estimate of the integral from $0$ to the largest root considered in the 
problem. This procedure transforms the integral of a rapidly oscillating integrand into parts within which the integrand is `slowly varying'. Of course, the function $\cos (y \omega_0 t - \phi(y) )$ has an infinite number of 
roots; consequently, an upper limit $\omega_c$ must be introduced in the estimation. Given the structure of $h(s \omega)$, in particular the factor $\exp (-\kappa \lvert s \omega / \omega_0 \rvert^c / c)$, if $\omega_c$ is carefully 
chosen, the significant part of the integral of Eq.~(\ref{eq:EQ0050}) is contained within the interval $(0, \omega_c]$. The threshold $\omega_c$ is defined herein as the $\omega$ value corresponding to $10^{-6}$ of the peak value 
of the envelope of the integrand, obtained with $y_e$ of Eq.~(\ref{eq:EQ0041}). Below thresholds of about $10^{-4}$, the results are insensitive to the choice of the fraction of the peak value used. Unfortunately, we did not find 
a way to apply the method of Hurwitz, Pfeiffer, and Zweifel \cite{hz,hpz} in this problem.

\subsection{\label{sec:SERe}The structure equations and re-assignment}

It was shown earlier that the integration of the Fourier transform of Eq.~(\ref{eq:EQ0046}) yields a real wavelet, see Eq.~(\ref{eq:EQ0050}). On the other hand, if the integration of $\hat{\psi}_{s \tau} (\omega)$ is restricted 
to positive $\omega$ values, one obtains a complex (holomorphic) wavelet. Exempting a factor of $2$ in the denominator of the overall factor on the right-hand side (rhs), the real part of the wavelet is still obtained via 
Eq.~(\ref{eq:EQ0050}), whereas the imaginary part is given by a similar expression, with a sine function replacing the cosine one in the integrand.
\begin{align} \label{eq:EQ0095}
\psi_{s \tau} (t) & = \frac{1}{\sqrt{2 \pi}} \int_{0}^{\infty} \hat{\psi}_{s \tau} (\omega) e^{i \omega t} d\omega\nonumber\\
&= \frac{k s^{\kappa \nu}}{2 \pi \omega^{\kappa \nu}_0} \int_{0}^{\infty} \cos \left( \omega t - \phi(\frac{s \omega}{\omega_0}) - \omega \tau \right) \, \exp \left( -\frac{\kappa s^c}{c\omega^c_0} \omega^c \right) \, \omega^{\kappa \nu - 1/2} d\omega\nonumber\\
&+ i \frac{k s^{\kappa \nu}}{2 \pi \omega^{\kappa \nu}_0} \int_{0}^{\infty} \sin \left( \omega t - \phi(\frac{s \omega}{\omega_0}) - \omega \tau \right) \, \exp \left(-\frac{\kappa s^c}{c\omega^c_0} \omega^c \right) \, \omega^{\kappa \nu - 1/2} d\omega\nonumber\\
&= \Re [ \psi_{s \tau} (t) ] + i \Im [ \psi_{s \tau} (t) ]
\end{align}
The operators $\Re$ and $\Im$ extract the real and the imaginary parts of the argument, respectively.

The absolute normalisation $k \in \mathbb{R}^+$ is now obtained via the expression
\begin{equation*}
k = \left( \frac{2 \kappa}{c} \right)^{\kappa \nu/c} \sqrt{\frac{2 \pi c}{\Gamma(\frac{2 \kappa \nu}{c})}} \, \, \, .
\end{equation*}
Taking Eqs.~(\ref{eq:EQ0015},\ref{eq:EQ0046},\ref{eq:EQ0047}) into account, the admissibility constant $c_{\psi}^2$ for the Reimann wavelets may be obtained by the formula
\begin{equation} \label{eq:EQ0055}
c_{\psi}^2 = \frac{2 \pi}{\omega_0} \left( \frac{2 \kappa}{c} \right)^{1/c} \frac{\Gamma \left( \frac{2 \kappa \nu - 1}{c} \right)}{\Gamma \left( \frac{2 \kappa \nu}{c} \right)} \, \, \, .
\end{equation}
In order that condition (\ref{eq:EQ0020}) be fulfilled, $\kappa \nu> 1/2$ and $c>0$.

\begin{figure}
\begin{center}
\includegraphics[width=15.5cm]{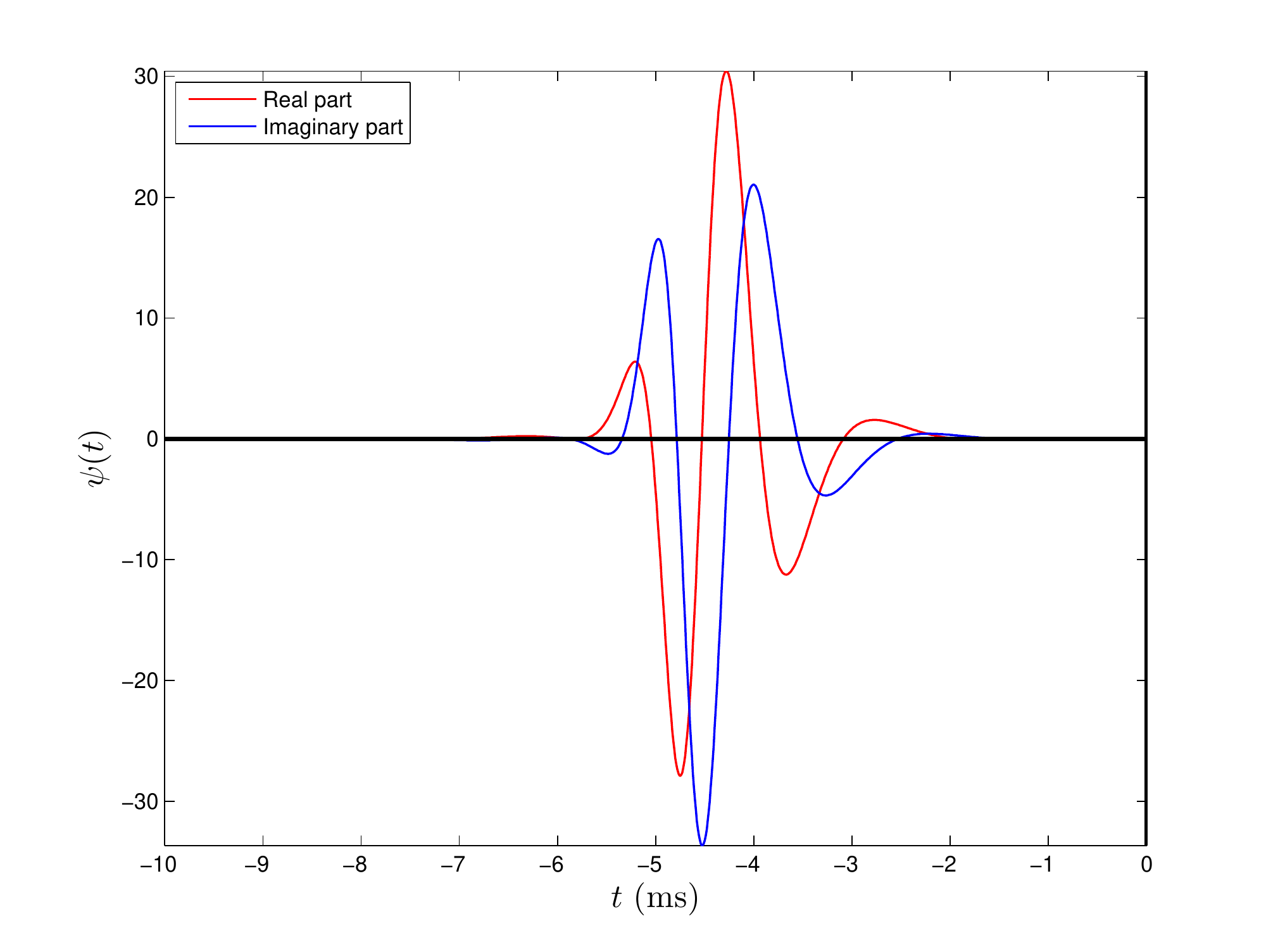}
\caption{\label{fig:wavelet}Plots of the real and imaginary parts of a wavelet in the time domain, obtained with Eq.~(\ref{eq:EQ0095}). The parameter values yielding these contributions are listed in Table \ref{tab:OptParVal}.}
\vspace{0.35cm}
\end{center}
\end{figure}

For complex wavelets, the transform of Eq.~(\ref{eq:EQ0011}) is also complex. The real and imaginary parts of one wavelet in the time domain are shown in Fig.~\ref{fig:wavelet}. The amplitude and the phase of the wavelet 
transform satisfy two differential equations, the `structure equations', introduced by Eqs.~(65,66) in Ref.~\cite{r1}.

A simple way to obtain the structure equations in a didactical manner is by making use of a harmonic signal $f (t) = f_s (t) = e^{i \omega_s t}$; the wavelet transform of this signal is (i.e., up to the normalisation factor 
$\sqrt{2 \pi}$) equal to the complex conjugate of the Fourier transform of the wavelet, evaluated at $\omega=\omega_s$:
\begin{equation*}
WT_{f_s}^\psi (s, \tau) = \sqrt{2 \pi} \hat{\psi}^*_{s \tau} (\omega_s) \, \, \, .
\end{equation*}
Using expressions (\ref{eq:EQ0046},\ref{eq:EQ0047}), one may put $WT_{f_s}^\psi (s, \tau)$ into the form
\begin{equation} \label{eq:EQ0100}
WT_{f_s}^\psi (s, \tau) = k \frac{s^{\kappa \nu}}{\omega^{\kappa \nu}_0} \, \exp \left( i \phi(\frac{s \omega_s}{\omega_0} ) + i \omega_s \tau \right) \, \exp \left( -\frac{\kappa \omega_s^c}{c \omega^c_0} s^c \right) \, \omega^{\kappa \nu - 1/2}_s \, \, \, .
\end{equation}
The differentiation of $WT_{f_s}^\psi (s, \tau)$ with respect to $s$ yields (after some trivial algebraical operations)
\begin{equation} \label{eq:EQ0110}
s \frac{\partial \ln WT_{f_s}^\psi (s, \tau)}{\partial s} = \kappa \nu + i \alpha - i \frac{\beta s}{\omega_0} \omega_s - \frac{\kappa s^c}{\omega^c_0} \omega_s^c \, \, \, .
\end{equation}
On the other hand, the differentiation of Eq.~(\ref{eq:EQ0100}) with respect to $\tau$ yields
\begin{equation} \label{eq:EQ0120}
\frac{\partial \ln WT_{f_s}^\psi (s, \tau)}{\partial \tau} = i \omega_s \, \, \, .
\end{equation}
To obtain the relation between the two derivatives, one simply replaces $\omega_s$ in Eq.~(\ref{eq:EQ0110}) by the result of Eq.~(\ref{eq:EQ0120}). The final expression, obtained in Ref.~\cite{r1} for arbitrary input, is
\begin{equation} \label{eq:EQ0130}
s \frac{\partial \ln WT_f^\psi (s, \tau)}{\partial s} = \kappa \nu + i \alpha - \frac{\beta s}{\omega_0} \frac{\partial \ln WT_f^\psi (s, \tau)}{\partial \tau} - \frac{\kappa s^c}{\omega^c_0} \left( \frac{1}{i} \frac{\partial \ln WT_f^\psi (s, \tau)}{\partial \tau} \right)^c \, \, \, .
\end{equation}

Invoking the polar representation for $WT_f^\psi (s, \tau) = r (s, \tau) e^{i \varphi (s, \tau)}$ and introducing the definitions
\begin{equation} \label{eq:EQ0140}
\frac{\partial \ln r (s, \tau)}{\partial s} = D_s^r \, \, \, ,
\frac{\partial \ln r (s, \tau)}{\partial \tau} = D_{\tau}^r \, \, \, ,
\frac{\partial \varphi (s, \tau)}{\partial s} = D_s^{\varphi} \, \, \, ,
\frac{\partial \varphi (s, \tau)}{\partial \tau} = D_{\tau}^{\varphi} \, \, \, ,
\end{equation}
one (after considering the real and imaginary parts of Eq.~(\ref{eq:EQ0130})) comes up with the set of equations:
\begin{equation*}
s D_s^r = \kappa \nu - \frac{\beta s}{\omega_0} D_{\tau}^r - \frac{\kappa s^c}{\omega^c_0} \Re \left[ \left( \frac{D_{\tau}^r}{i} + D_{\tau}^{\varphi} \right) ^c \right] 
\end{equation*}
and
\begin{equation*}
s D_s^{\varphi} = \alpha - \frac{\beta s}{\omega_0} D_{\tau}^{\varphi} - \frac{\kappa s^c}{\omega^c_0} \Im \left[ \left( \frac{D_{\tau}^r}{i} + D_{\tau}^{\varphi} \right) ^c \right] \, \, \, .
\end{equation*}
These are the structure equations. In the second part of this subsection, it will become evident that the two derivatives of the phase $\varphi (s, \tau)$, i.e., $D_s^{\varphi}$ and $D_{\tau}^{\varphi}$, are the interesting 
quantities. The derivatives $D_s^r$ and $D_{\tau}^r$ are easily obtained; due to the branching of the phase, there are no unique solutions for $D_s^{\varphi}$ and $D_{\tau}^{\varphi}$ in the general case $c \in \mathbb{R}^+$. 
To overcome the uniqueness problem, Ref.~\cite{r1} performs an expansion of the last term on the rhs of Eq.~(\ref{eq:EQ0110}) around $s \omega_s / \omega_0=1$; this approximation is performed in the frequency domain. In the 
present paper, the generality of the approach will be dropped, in favour of the uniqueness of the solution; as a result, the parameter $c$ will be fixed to $1$. This approach has the obvious advantage that no approximation of 
Eq.~(\ref{eq:EQ0110}) is necessary and unique solutions for $D_s^{\varphi}$ and $D_{\tau}^{\varphi}$ are obtained.

Up to now, general formulae have been given, i.e., applicable for arbitrary $\nu$ and $c$. This might be useful to other studies, in case that different assumptions on the parameters of the Reimann wavelets are made. From now on, 
it will be assumed that $\nu=c=1$. As a result, Eq.~(\ref{eq:EQ0110}) yields two simpler structure equations:
\begin{equation} \label{eq:EQ0160}
D_{\tau}^{\varphi} = \frac{\omega_0}{s} - \frac{1}{\kappa} ( \beta D_{\tau}^r + \omega_0 D_s^r )
\end{equation}
and
\begin{equation} \label{eq:EQ0170}
s D_s^{\varphi} = \alpha - \frac{s}{\omega_0} ( \beta D_{\tau}^{\varphi} - \kappa D_{\tau}^r ) \, \, \, .
\end{equation}
Steps towards incorporating the re-assignment method into the analysis framework can be taken now.

The method of re-assignment for the analysis of non-stationary signals was set forth by Kodera, Gendrin, and de Villedary \cite{kgv} in 1978. In their recent paper \cite{gm}, Gardner and Magnasco made the point that the 
time-frequency representations of the auditory channel are sparse, i.e., that most of the neurons are inactive for most of the time, and advance the thesis that these neurons perform some kind of time-frequency analysis of the 
phase of acoustical signals; for instance, time derivatives of the phase may be obtained from single fibres of the auditory nerve, whereas frequency derivatives may be `evaluated' from tonotopically organised fibres. Seen from 
the point of view of the implementation of the re-assignment method, the time derivative of the phase of the Gabor transform $\chi(\omega,t)$ (see Eq.~(1) of Ref.~\cite{gm}), which may be considered analogous to the wavelet 
transform of the present work, defines the instantaneous angular frequency $\tilde{\omega}$ (denoted by $\omega_{ins}$ therein), whereas the frequency derivative enters the definition of the instantaneous time $\tilde{t}$ 
(denoted by $t_{ins}$ therein):
\begin{equation} \label{eq:EQ0200}
\tilde{\omega} = \frac{\partial \varphi (\omega,t)}{\partial t}
\end{equation}
and
\begin{equation} \label{eq:EQ0210}
\tilde{t} = t - \frac{\partial \varphi (\omega,t)}{\partial \omega} \, \, \, .
\end{equation}
Via this transformation, points of the ($\omega$,$t$) plane are mapped onto ($\tilde{\omega}$,$\tilde{t}$). Various weights may be used in the mapping, e.g., equal weights (the Lebesgue measure) or the spectrogram 
$\lvert \chi(\omega,t) \rvert^2$ (thus, creating the re-assigned spectrogram). Reference \cite{gm} also offers another alternative, namely re-assigning $\chi(\omega,t)$ by histogramming ($\tilde{\omega}$,$\tilde{t}$) with 
the complex weight factor $\chi(\omega,t) e^{i(\tilde{\omega}+\omega)(\tilde{t}-t)/2}$. Since the phase information is included in the mapping, this last choice appears promising in signal processing.

One of the main points of Ref.~\cite{gm} is that the noise and the (common) signals generate different patterns on the ($\tilde{\omega}$,$\tilde{t}$) plane; as a result, noise-suppression techniques may be developed, retrieving 
the useful signal from noisy data. In Ref.~\cite{gm}, the method has been successfully applied to a number of simple signals (harmonic, clicks, sweeps, and chirps).

The details of the involvement of the re-assignment method in this analysis framework were recently worked out \cite{r3}. A short summary of the important steps is given next, placing the emphasis on two issues: a) the assessment 
of the importance of the wavelet coefficients obtained via Eq.~(\ref{eq:EQ0011}) and b) the suppression of the noise present in the original signal.

The modification of Eqs.~(\ref{eq:EQ0200},\ref{eq:EQ0210}) in order to match the WBA involves two associations: the angular frequency $\omega$ is related to the inverse of the scaling factor $s$ and the time $t$ to the time-shift 
factor $\tau$. One thus obtains the equivalent forms of Eqs.~(\ref{eq:EQ0200},\ref{eq:EQ0210}).
\begin{equation*}
\frac{\omega_0}{\tilde{s}} = D_{\tau}^{\varphi}
\end{equation*}
and
\begin{equation*}
\tilde{\tau} = \tau + \frac{s^2}{\omega_0} D_s^{\varphi} \, \, \, .
\end{equation*}

Regarding the weights to be applied in the mapping ($s$,$\tau$) $\rightarrow$ ($\tilde{s}$,$\tilde{\tau}$), it is recommended, analogously to the FBA \cite{gm}, to histogram ($\tilde{s}$,$\tilde{\tau}$) with the complex weight 
factor obtained after the multiplication of the wavelet transform of the signal by $\exp (i \omega_0 ( \tilde{s}^{-1} + s^{-1} ) (\tilde{\tau}-\tau) / 2 )$; according to Ref.~\cite{gm}, such a choice retains the information on 
the phase of the wavelet transform and should be taken as the counterpart of the complex re-assigned STFT, discussed in Ref.~\cite{gm}, p.~6097.

\section{\label{sec:Results}Results}

\subsection{\label{sec:Processing}Settings in the data processing}

Each audio file is processed as follows. A window equivalent to $N_M$ sound-wave measurements is run over the input data, advancing by $(1-d) N_M$ measurements after each processing step; the quantity $d \in (0,1)$ determines the 
overlap between successive windows. The quality of the reconstructed signal is expected to improve with increasing overlap, which (inevitably) stretches the runtime load; it appears that an overlap between $50$ and $75 \%$ 
achieves a good compromise between quality and speed. At each processing step, the $N_M$ measurements are processed (via the wavelet transformations of Eqs.~(\ref{eq:EQ0011},\ref{eq:EQ0013})), yielding `predictions' for the 
$(1-d) N_M$ central elements of the window (reconstructed values); these predictions constitute the output of the processing at the given window position. The window then advances to the next input-data segment (shift by $(1-d) N_M$ 
measurements); the process continues until the data is exhausted. Evidently, following this procedure, all data (save for $d N_M / 2$ values at the beginning and an equal amount of values at the end of the audio file) are 
reconstructed.

The range of the time-shift factor $\tau$, used in the wavelet transformations, depends on the characteristics of the specific wavelet being employed, namely on the range of its significant values in the time domain. It appears 
reasonable to associate the $\tau$ domain ($\tau_R$) with $N_M$. In this work, $\tau_R$ was set to an equivalent length of $8 N_M$ measurements ($\tau_R=8 N_M \Delta t$, where $\Delta t$ denotes the sampling interval); good 
results were also obtained with a smaller window, spanning $4 N_M$ measurements, which is more appropriate (significantly less time-consuming) for real-time applications~\footnote{As the mother wavelet will not have positive 
support (see Subsection \ref{sec:Parameters}), the midpoints of the intervals $\tau_R$ and $N_M \Delta t$ do not coincide.}. The step in $\tau$ is denoted by $\Delta \tau$.

In the chromatic musical scale, one octave comprises five tone intervals ($C$ to $D$, $D$ to $E$, $F$ to $G$, $G$ to $A$, and $A$ to $B$) and two semitone ones ($B$ to $C$ and $E$ to $F$), hence, twelve semitone intervals in 
total~\footnote{Used herein is the naming standard which is followed in the United States, Canada, the United Kingdom, and Ireland, rather than the one which several (non-Romance) countries of Continental Europe, i.e., the Germanic 
countries, as well as Russia, Poland, and Scandinavia have adopted. According to the latter, the note $B$ of the former is named $H$, whereas $B\flat$ is named $B$.}. The resolution efficiency of the human ear is believed to be 
about a half semitone. Given that the same notes (e.g., $A$'s) in successive octaves correspond to a ratio of $2$ in frequency, each semitone interval involves a frequency ratio of $\sqrt[12]{2} \approx 1.06$. In Subsection 
\ref{sec:WaveletTimeDomain}, the extreme values of the angular frequency of acoustical signals were set to $\omega_{\rm max}=2 \pi 20\,000$ rad/s and $\omega_{\rm min}=2 \pi 60$ rad/s; therefore, the available frequency range 
spans $8.38$ octaves (i.e., the $\gamma$ value divided by $\ln 2$) or, equivalently, about $100.57$ semitone intervals. Therefore, in order to set the step size in $s$ to about a half semitone, one simply divides the [$0$,$1$] 
interval of $x$ into $200$ (equal) segments. (Evidently, semitone and tone resolutions correspond to about $100$ and $50$ segments, respectively.) The resolution in $s$ is denoted by $\Delta s$.

The processing of acoustical data depends on the values of the quantities $N_M$, $d$, $\Delta s$, and $\Delta \tau$. Despite the fact that these quantities are parameters, they are rather considered as `settings' herein; the 
term `parameters' is reserved for the quantities associated with the wavelet form (see Subsection \ref{sec:WaveletTimeDomain}).
\begin{itemize}
\item One standard choice for $N_M$ (for manufacturers of sound-processing equipment) is $128$ (power of $2$, enabling the application of fast DFT algorithms). In case that $\tau_R$ is linked to $N_M$, the memory and runtime 
requirements in WBAs are expected to rise as $N_M^2$. This is due to the fact that doubling the window size leads to (more than) quadrupling the dimension of the important arrays used in the data processing, which (as operations 
are performed on the elements of these arrays) inevitably results in the increase of the runtime by the same factor. To keep the memory and runtime requirements at minimal levels, we therefore focuss on $N_M=128$ herein, though 
results with $N_M=256$ were also obtained for the sake of comparison.
\item Representative values for the overlap $d$ are: $50$ and $75 \%$.
\item Reasonable values for $\Delta s$ are: a half semitone, one semitone, and one tone.
\item Reasonable values for $\Delta \tau$ are: $2$, $4$, and $8 \Delta t$.
\end{itemize}
Concerning $\Delta s$ and $\Delta \tau$, the quality of the reconstruction deteriorates from left to right in the scheme above; at the same time, the memory and runtime requirements become less demanding from left to right. 
Evidently, the choice of the settings involves a trade-off between the quality of the reproduction, the memory requirements, and the runtime demands; as such, it must reflect the placement of the emphasis in a study. For 
instance, when investigating the optimal parameter values of the Reimann wavelets, memory and speed are not relevant. On the contrary, in real-time processing of acoustical data, memory and speed are paramount; in order to 
obtain fast results in such applications (and avoid undue delays), the expectations on quality must be somewhat curtailed.

\subsection{\label{sec:Parameters}Optimal parameter values}

For the determination of optimal parameter values, it makes sense to use high resolution in the settings pertaining to the reconstruction of the data. The values of $N_M$, $d$, $\Delta s$, and $\Delta \tau$ were: $128$, $75 \%$, 
a half semitone, and $4 \Delta t$, respectively; additionally, $\tau_R=8 N_M \Delta t$.

We will now discuss the issue of causality in relation to the present analysis framework. Causality dictates that the output of the processing at time instant $t_0$ may depend only on information acquired at former times, i.e., 
at $t<t_0$. On the other hand, the present application involves operations on $N_M$ sound-wave measurements at each window position. Using, for the sake of the example, the value of $20$ kHz for the sampling frequency, $N_M=128$, 
and $d=75 \%$, the data within the last $6.4$ ms, prior to the `current' time, are submitted for analysis and comprise the output for time instants between $-(1-d/2) N_M \Delta t=-4.0$ and $-d N_M \Delta t/2=-2.4$ ms, i.e., a few 
ms in the past, prior to the `current' time. The window then advances by $(1-d) N_M \Delta t=1.6$ ms. The generation of the output obeys a cycle performed $625$ times per second~\footnote{In practice, the sound-wave measurements 
are `continuously' captured, filling up an array, which is submitted for processing each time it is filled. The last $d N_M$ elements of this array are retained, becoming the first part of the subsequent array of $N_M$ measurements.}. 
It is evident that the minimal delay (difference between the time at which a sound-wave measurement is captured and that at which the corresponding output becomes available) is equal to $(1-d/2) N_M \Delta t=4.0$ ms (at best). 
(To achieve minimal delay values, $d$ is chosen as large as possible. However, this has drawbacks in real-time applications, as it stretches the runtime load.) In any case, the point is that regardless of whether the analysis of 
an acoustical signal is Fourier- or wavelet-based, $N_M$ measurements are \emph{simultaneously} submitted for processing, the central $(1-d) N_M$ values of which comprise the output at the particular window position. As a result, 
it is not clear whether and how causality enters the analysis, and whether the wavelets need to be causal. The issue of causality has not been addressed in Refs.~\cite{r1,r2}.

In spite of the fact that the reproduction of the data is hardly affected, we will assume the cautious attitude of considering only causal wavelets herein, i.e., mother wavelets with suppressed positive support~\footnote{Slightly 
better results are obtained with non-causal wavelets; this is not surprising as, in comparison to free fits, constrained fits are bound to yield inferior results.}. After scanning the parameter space, it appeared that one good 
starting point, to be used as `initial guess' in the optimisation scheme, was: $\alpha= \pi$, $\beta=8.5 \pi$, $\phi_m = -2 \pi$, $\kappa=8$ (as mentioned earlier, the parameters $\nu$ and $c$ are fixed to $1$). On the way to 
extract optimal parameter values from a data set, regardless of the domain of application, three are the main issues: a) the quantification of a qualitative concept such as the `goodness of the description of the data', b) 
the method which is employed for obtaining the optimal parameter values, and c) the data on the basis of which the determination is made. These three points will be addressed next.

The statistical measure of the quality of the data description is the linear (Pearson's) correlation coefficient $\rho$ between the input and the output values. For arrays obeying an ideal linear relation, $\rho = 1$ (or 
$\rho = -1$ in case of an ideal anti-correlation, which is not of relevance herein); absence of correlation results in $\rho = 0$. The linear correlation coefficient is a good measure of the quality of the data description in 
case of perfect (i.e., noise-free) input; the effects of the noise will be touched upon in Subsection \ref{sec:Noise}.

We now describe the method of extracting the optimal parameter values. Commencing from the parameters pertaining to the initial guess, a three-step approach was followed. a) Each parameter was set to three values: the one 
corresponding to the aforementioned initial vector in the parameter space (central value) and the two values defined at $\pm 10 \%$ of the central value. In case that the $\rho$ value was largest for the central element, a 
quadratic fit yielded the candidate optimal value, which was either accepted (if the $\rho$ value at the new setting exceeded the one which had been previously obtained for the central element) or rejected (in which case, the 
central element was used as the optimal setting). If, on the other hand, the central value of the parameter did not yield the largest $\rho$ value, the parameter was varied in the direction of maximising $\rho$ (always using 
the maximal step of $10 \%$ of the starting value), until three values were found for which the central one yielded the largest $\rho$ (in other words, the procedure was interrupted at the moment when the $\rho$ maximum was 
bracketed); a new quadratic fit determined the candidate optimal value, which was either accepted or rejected (by applying the aforementioned criterion). The parameters were treated serially, in the order: $\beta$, $\phi_m$, 
$\alpha$, and $\kappa$; this order was obtained on the basis of decreasing importance in regard to changes induced on the resulting wavelet (determined via a simple variational analysis around the parameter-space point 
corresponding to the initial guess). b) Step (a) was repeated, with $3 \%$ (instead of $10 \%$) steps for each parameter, and using the final result of step (a) as the initial parameter vector. c) Step (a) was repeated, with 
$1 \%$ steps for each parameter, and using the final result of step (b) as the initial parameter vector. All the steps above involved only causal wavelets; if a non-causal wavelet was suggested at any point in the procedure 
just outlined, that wavelet was replaced by a close-by (within the `current' interval of variation of the particular parameter) causal neighbour (randomly selected).

The important results for the fitted parameters and for $\rho$ are shown in Table \ref{tab:OptParVal}. One notices that, during the process of obtaining these values, the overall improvement (increase in $\rho$) is insignificant. 
From now on, the final result of Table \ref{tab:OptParVal} will be referred to as `standard'; the figures in this study have been obtained on the basis of this parameter vector.

One point ought to be mentioned. The correlations among the model parameters are large; these correlations may be inherent (i.e., relating to the form of the Reimann wavelets) and/or method-related (i.e., induced by the technique 
used to extract the important information, namely by the application of transformations whose product is the identity transformation).

The audio file, on which the optimal parameter values have been obtained, comprises $5$-s patches of harmonic signals of unit amplitude, corresponding to successive $A$'s of the chromatic musical scale, from $A_2$ ($110$ Hz) to 
$A_7$ ($3\,520$ Hz). To ensure the continuity of the first derivative of the signal, the individual files were `patched together' at time instants corresponding to the maxima (where the first derivative of the signal with respect 
to time vanishes). The sampling frequency was equal to $28\,160$ Hz. The resulting $30$-s file contains $844\,800$ one-channel ($16$-bit) sound-wave measurements. The input frequencies were treated on equal footing; one could 
also introduce weights, to account for the importance of the various frequencies in human-related acoustics.

\subsection{\label{sec:Harmonic}Reconstruction of harmonic signals}

\begin{figure}
\begin{center}
\includegraphics[width=15.5cm]{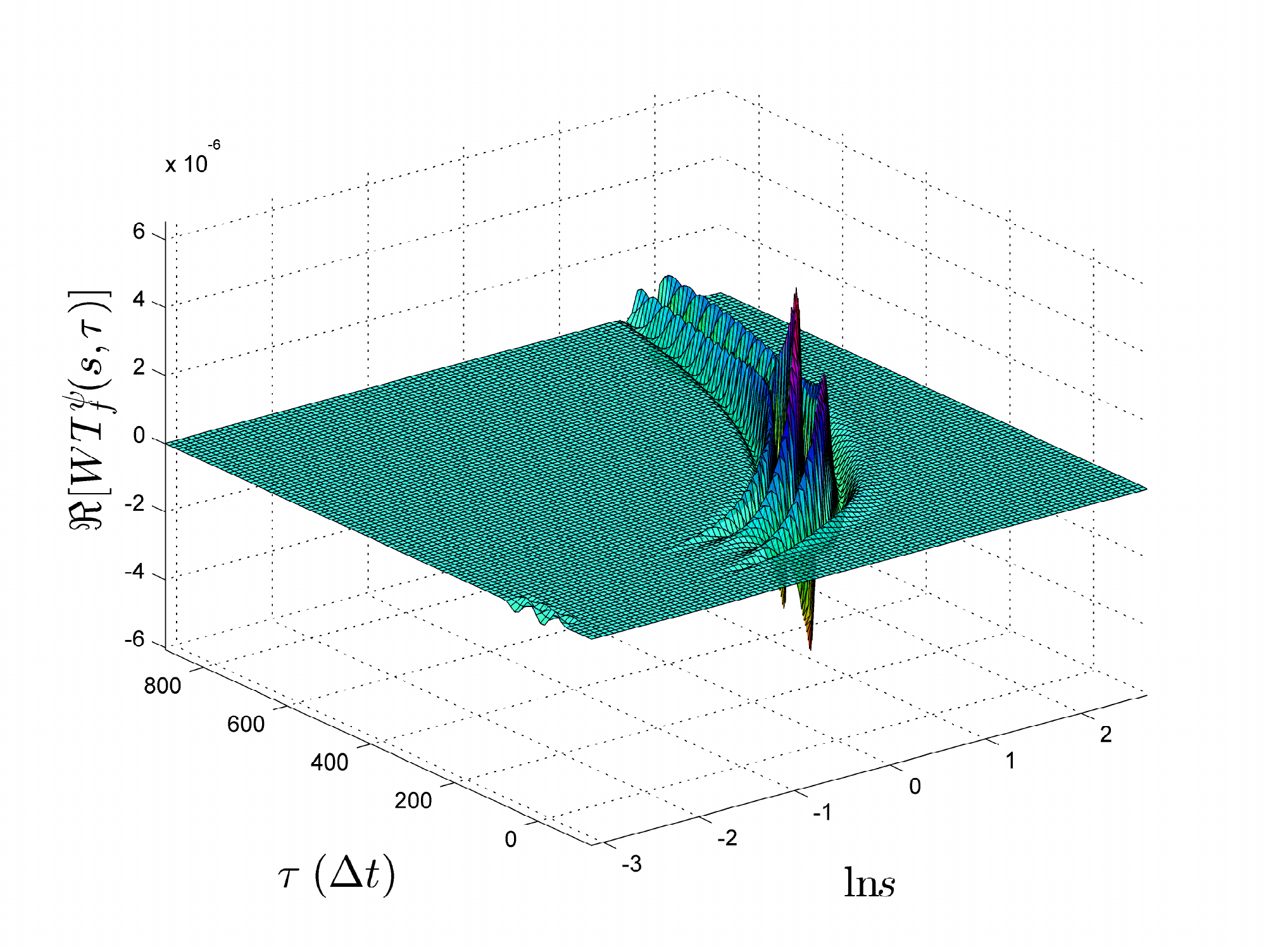}
\caption{\label{fig:WaveletTransformRl}The real part of the wavelet transform at $440$ Hz as a function of $\ln s$ and $\tau$. The mother wavelet, obtained with the standard parameter values (see Table \ref{tab:OptParVal}), has 
been used.}
\vspace{0.35cm}
\end{center}
\end{figure}

\begin{figure}
\begin{center}
\includegraphics[width=15.5cm]{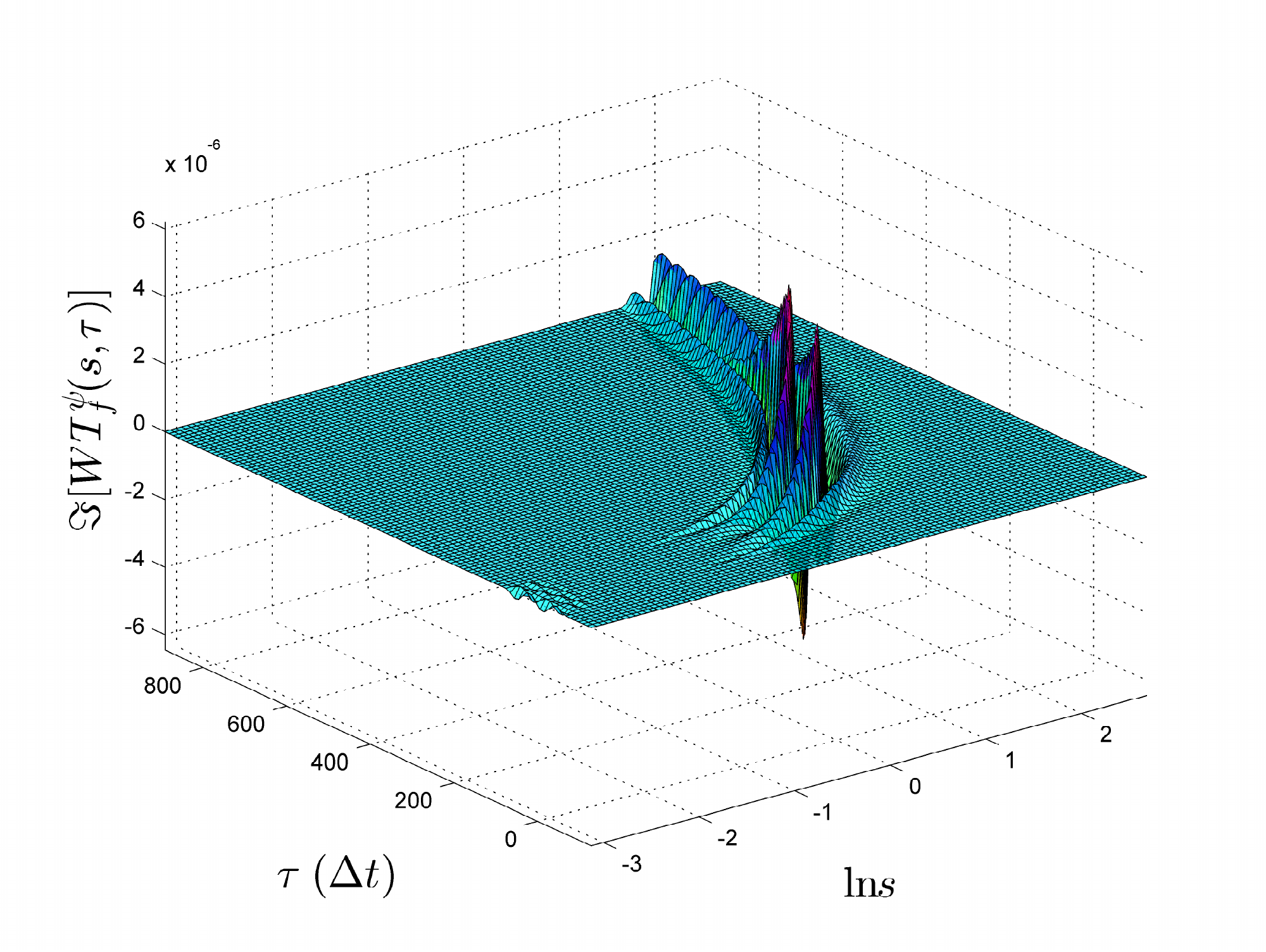}
\caption{\label{fig:WaveletTransformIg}The imaginary part of the wavelet transform at $440$ Hz as a function of $\ln s$ and $\tau$. The mother wavelet, obtained with the standard parameter values (see Table \ref{tab:OptParVal}), 
has been used.}
\vspace{0.35cm}
\end{center}
\end{figure}

\begin{figure}
\begin{center}
\includegraphics[width=15.5cm]{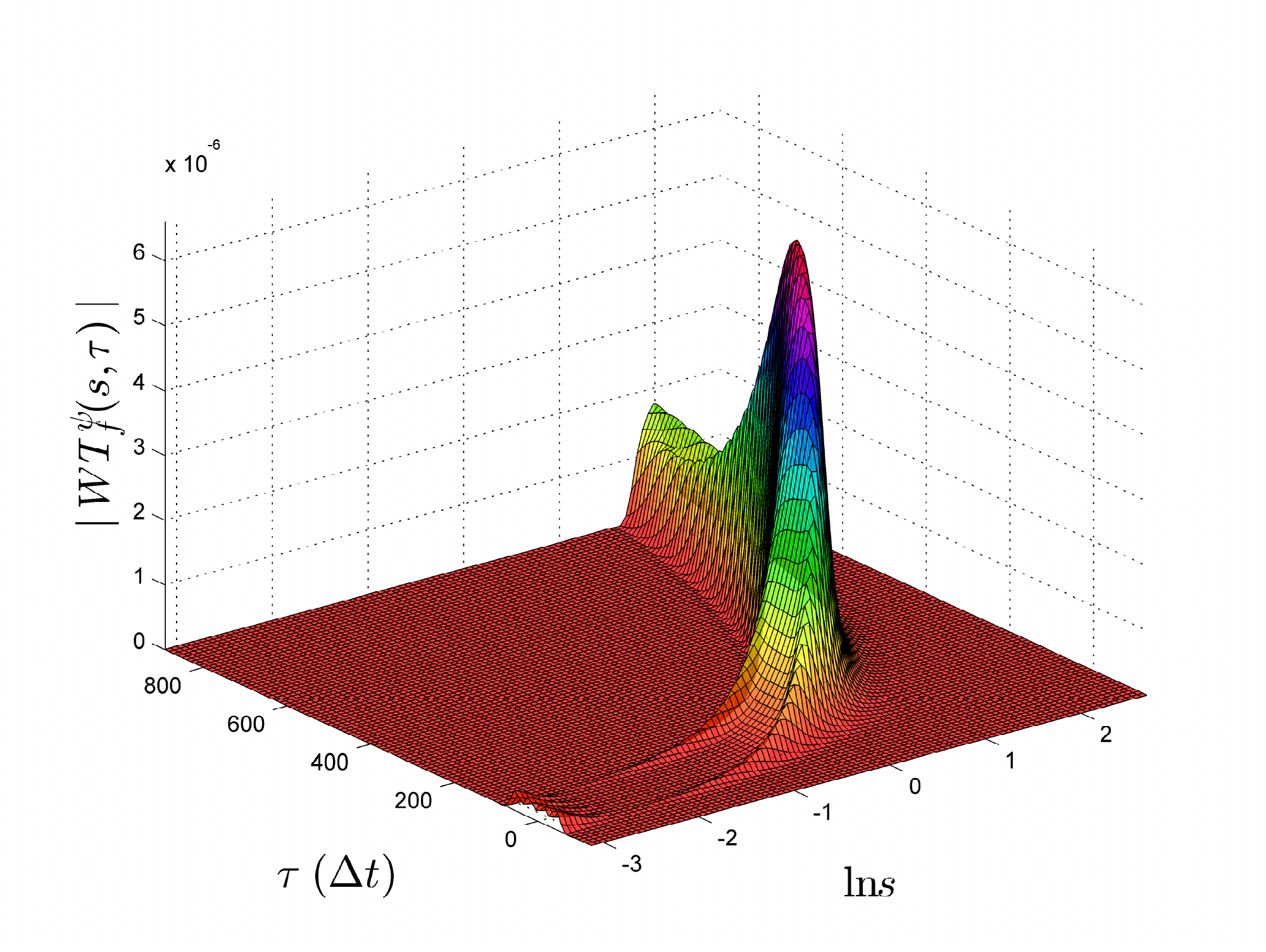}
\caption{\label{fig:WaveletTransform}The modulus of the wavelet transform at $440$ Hz as a function of $\ln s$ and $\tau$. The mother wavelet, obtained with the standard parameter values (see Table \ref{tab:OptParVal}), has been 
used. The peak value of the modulus of the wavelet transform appears, as expected, at $s=2$.}
\vspace{0.35cm}
\end{center}
\end{figure}

In the Scientific pitch notation \cite{spn}, $A_4$ denotes the note $A$ above middle $C$; the usual choice when tuning musical instruments is to set $A_4$ to $440$ Hz. A $5$-s harmonic signal at this frequency was generated and 
processed using the direct and inverse wavelet transformations given in Eqs.~(\ref{eq:EQ0011},\ref{eq:EQ0013}). The linear correlation coefficient between the input and the reconstructed data was $\rho \approx 0.999860$. The real 
and imaginary parts of the wavelet transform (average maps over the time span of the input data) are shown in Figs.~\ref{fig:WaveletTransformRl} and \ref{fig:WaveletTransformIg}, respectively; the corresponding modulus of the 
wavelet transform is shown in Fig.~\ref{fig:WaveletTransform}.

\begin{figure}
\begin{center}
\includegraphics[width=15.5cm]{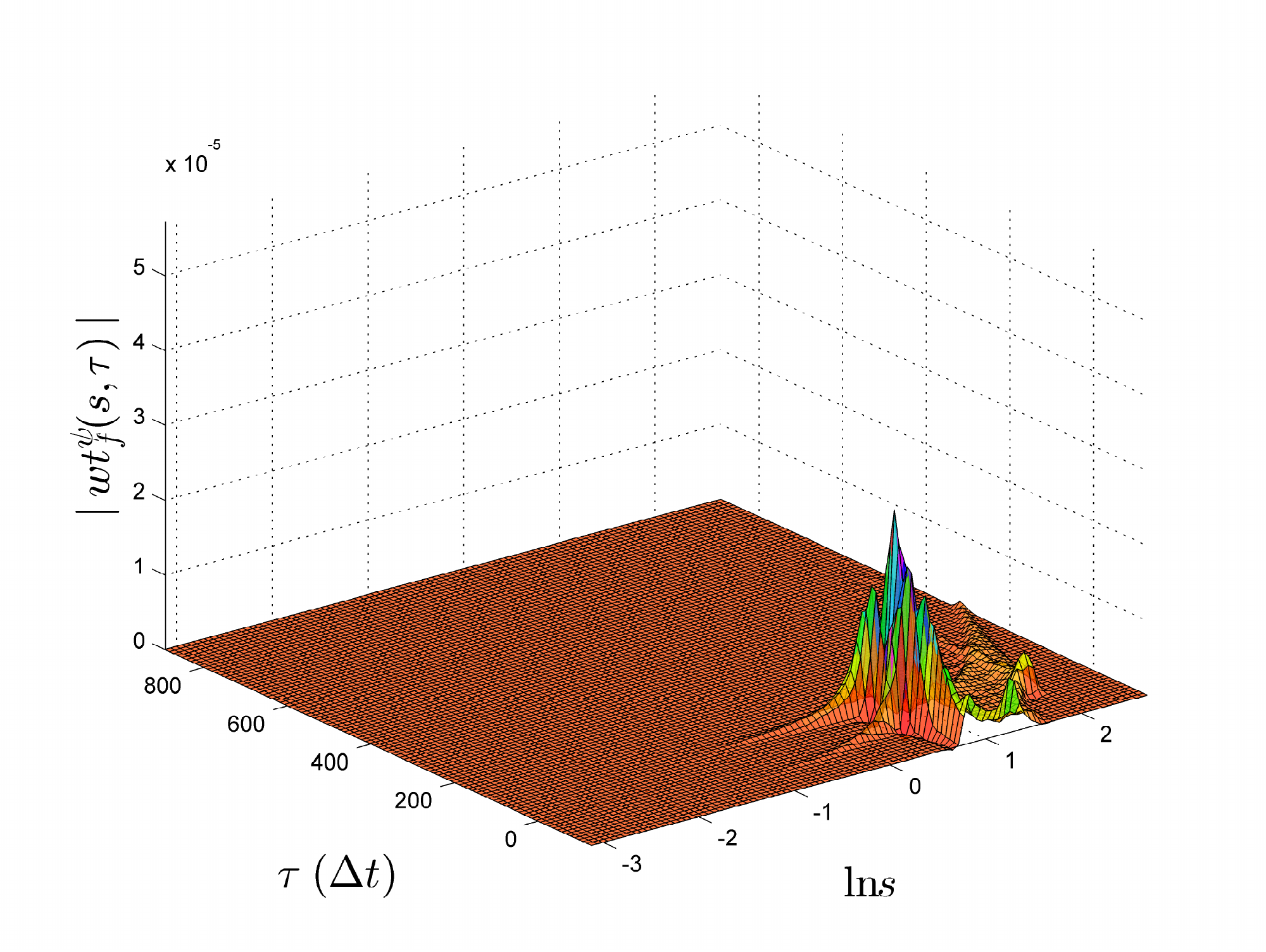}
\caption{\label{fig:ReWaveletTransform}The modulus of the complex re-assigned wavelet transform at $440$ Hz as a function of $\ln s$ and $\tau$. The mother wavelet, obtained with the standard parameter values (see Table 
\ref{tab:OptParVal}), has been used. The peak value of the modulus of the wavelet transform appears, as expected, at $s=2$.}
\vspace{0.35cm}
\end{center}
\end{figure}

At this point, the obvious question is whether the reproduction of these data improves after invoking the theoretical framework of the structure equations, enhanced with the re-assignment method, as presented in Subsection 
\ref{sec:SERe}. To this end, the complex re-assigned wavelet transform was obtained (see Fig.~\ref{fig:ReWaveletTransform}) and used in order to single out the dominant contributions when performing the inverse wavelet 
transformation. By doing this, without any processing of the re-assigned wavelet transform, a $\rho$ value is obtained, differing from $1$ by no more than $10^{-6}$. No audible difference between the original and the reconstructed 
audio files could be heard. As a result, the involvement of the mathematical framework of Subsection \ref{sec:SERe} yields encouraging results for noise-free input, even with a minor involvement of the re-assigned wavelet 
transform, such as the determination of the important~\footnote{The `importance threshold' was set to $10^{-12}$, several orders of magnitude smaller than the typical peak values in 
Fig.~\ref{fig:ReWaveletTransform}.} components in the signal.

To investigate the variation of the quality of the reconstruction with the frequency, a number of harmonic signals were generated at various frequencies, corresponding to multiples of $110$ Hz (powers of $2$, spanning six octaves): 
$110$, $220$, $440$, $880$, $1\,760$, $3\,520$, and $7\,040$ Hz, representing $A_2$ to $A_8$. Each audio file was $5$ s long. Given in Table \ref{tab:OrglAndRcstvsFrequency} are the $\rho$ values between the input and the 
reconstructed data. The quality of the reproduction is good over the frequency domain chosen, save for the highest frequency, where some deterioration in the quality of the processed audio file was observed; it should be borne 
in mind that the Nyquist frequency is equal to $14\,080$ Hz. Finally, the quality of the reconstruction was also investigated at $80$ Hz. As the lowest frequency (which may be resolved in a FBA using windows of $128$ measurements 
and a sampling frequency of $28\,160$ Hz) is $220$ Hz, it cannot but be regarded as a positive sign that the quality of the reconstruction (when processing the signals via the wavelet transformations) did not deteriorate at $80$ 
Hz: the $\rho$ value, obtained at $80$ Hz, was $0.999036$. One may conclude that a WBA performs well in the reproduction of the low-frequency components of acoustical signals.

\subsection{\label{sec:Noise}Noise}

By no means do we intend to place the emphasis on the subject of the noise reduction/suppression in the present paper. Our intention is only to demonstrate that the theoretical background of the work (as outlined in Section 
\ref{sec:Method}) constitutes a promising basis for further research and development. A more detailed study of the effects of the noise and on the development of dedicated algorithms (within this analysis framework) for its 
efficient reduction/suppression is currently under planning; that study will address the subject in an organised manner.

So far, perfect (noise-free) input data have been analysed. In the general case of noisy input, it is not trivial to define a measure of goodness of the processing. As the amount of the noise in the input data is, generally 
speaking, not straightforward to determine, generated data comprise the best means to study reliably the noise-related effects; to this end, we will use the harmonic signal at $440$ Hz of the previous subsection, add white noise 
to it, and investigate whether the processing of the resulting data may be performed in such a way that the noise be reduced. As the original input is noise-free, it represents the optimal output of the denoising of the input 
data; it is this array which the reconstructed data must be compared to. Three data arrays need to be compared: the original noise-free input data, the data after noise has been added, and the data obtained after processing the 
noisy input. Evidently, if the processing is successful, the reconstructed data should come out as close to the original data as possible; the larger the $\rho$ value between these two data arrays, the more efficient the 
processing is in terms of the noise suppression. If the noise is altogether eliminated, the $\rho$ value should come out as large as the entries of Table \ref{tab:OrglAndRcstvsFrequency}, where no noise was present in the input.

\begin{figure}
\begin{center}
\includegraphics[width=15.5cm]{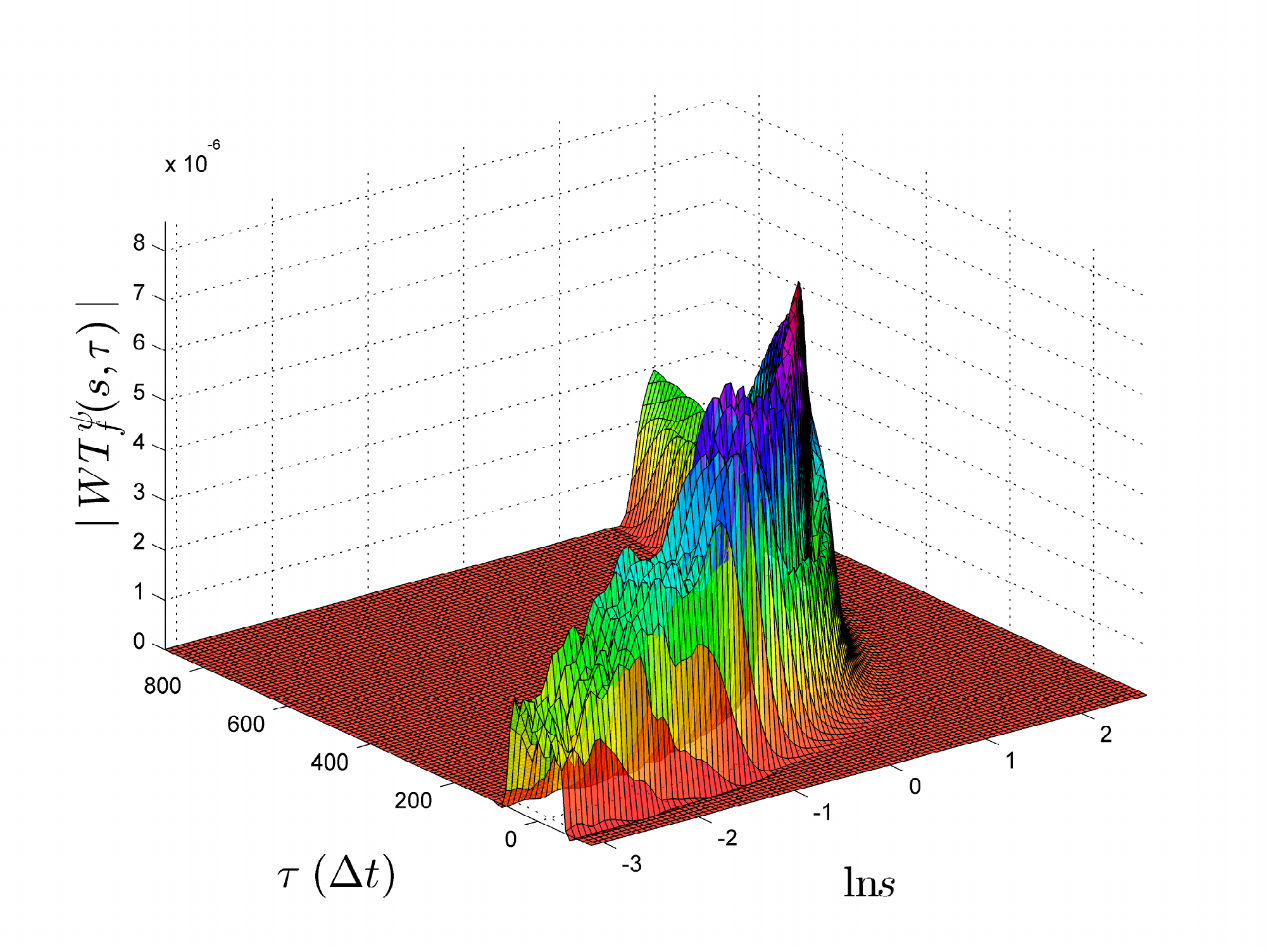}
\caption{\label{fig:WaveletTransformWith5PercentNoise}The modulus of the wavelet transform at $440$ Hz as a function of $\ln s$ and $\tau$ after $5 \%$ white noise was added to the original harmonic signal. The difference between 
this plot and the one shown in Fig.~\ref{fig:WaveletTransform} is the effect of the noise. The mother wavelet, obtained with the standard parameter values (see Table \ref{tab:OptParVal}), has been used.}
\vspace{0.35cm}
\end{center}
\end{figure}

$140\,800$ random numbers, following the normal N($0$,$1$) distribution, were generated and added onto the $5$-s, $440$-Hz harmonic signal of unit amplitude of Subsection \ref{sec:Harmonic}. The noise level in the final data was 
set equal to $5 \%$. This is a large amount of noise, as by listening to a number of simulated data, one concludes that amounts of noise even below $0.5 \%$ generate audible effects; the human ear appears to be an efficient noise 
detector. For the sake of completeness, the wavelet transform in the case of the noisy input is also given, see Fig.~\ref{fig:WaveletTransformWith5PercentNoise}; by comparing Figs.~\ref{fig:WaveletTransform} and 
\ref{fig:WaveletTransformWith5PercentNoise}, one obtains an impression of the noise distribution in the plot of the wavelet transform. Evidently, the most salient feature in Fig.~\ref{fig:WaveletTransformWith5PercentNoise} (in 
comparison to Fig.~\ref{fig:WaveletTransform}) is the high-frequency tail of the distribution; this tail must be eliminated in the processing.

In any case, the resulting data were analysed with three methods:
\begin{itemize}
\item Only the direct and the inverse wavelet transformations of Eqs.~(\ref{eq:EQ0011},\ref{eq:EQ0013}) were performed.
\item The complex re-assigned wavelet transform was used in order to establish the important contributions in the application of the inverse wavelet transform of Eq.~(\ref{eq:EQ0013}).
\item The complex re-assigned wavelet transform was processed (before being used in order to establish the important contributions in the application of the inverse wavelet transform of Eq.~(\ref{eq:EQ0013})). The adopted 
processing was very simple. In comparison to the signal, the noise appears to be distributed differently on the re-assigned ($\tilde{s}$,$\tilde{\tau}$) plot \cite{gm}. A simple cut in the connectivity plot was made use of. The 
connectivity plot is a map containing each pixel's number of important neighbours; each pixel in this plot contains values between $0$ and $8$ (in two dimensions). As noise tends to be distributed in the ($\tilde{s}$,$\tilde{\tau}$) 
plot, noise-related pixels are expected to have fewer important neighbours (than signal-related ones). The minimal number of neighbours for acceptable pixels is arbitrary; in the present study, each such pixel was assumed to 
have at least four important neighbours.
\end{itemize}

\begin{figure}
\begin{center}
\includegraphics[width=15.5cm]{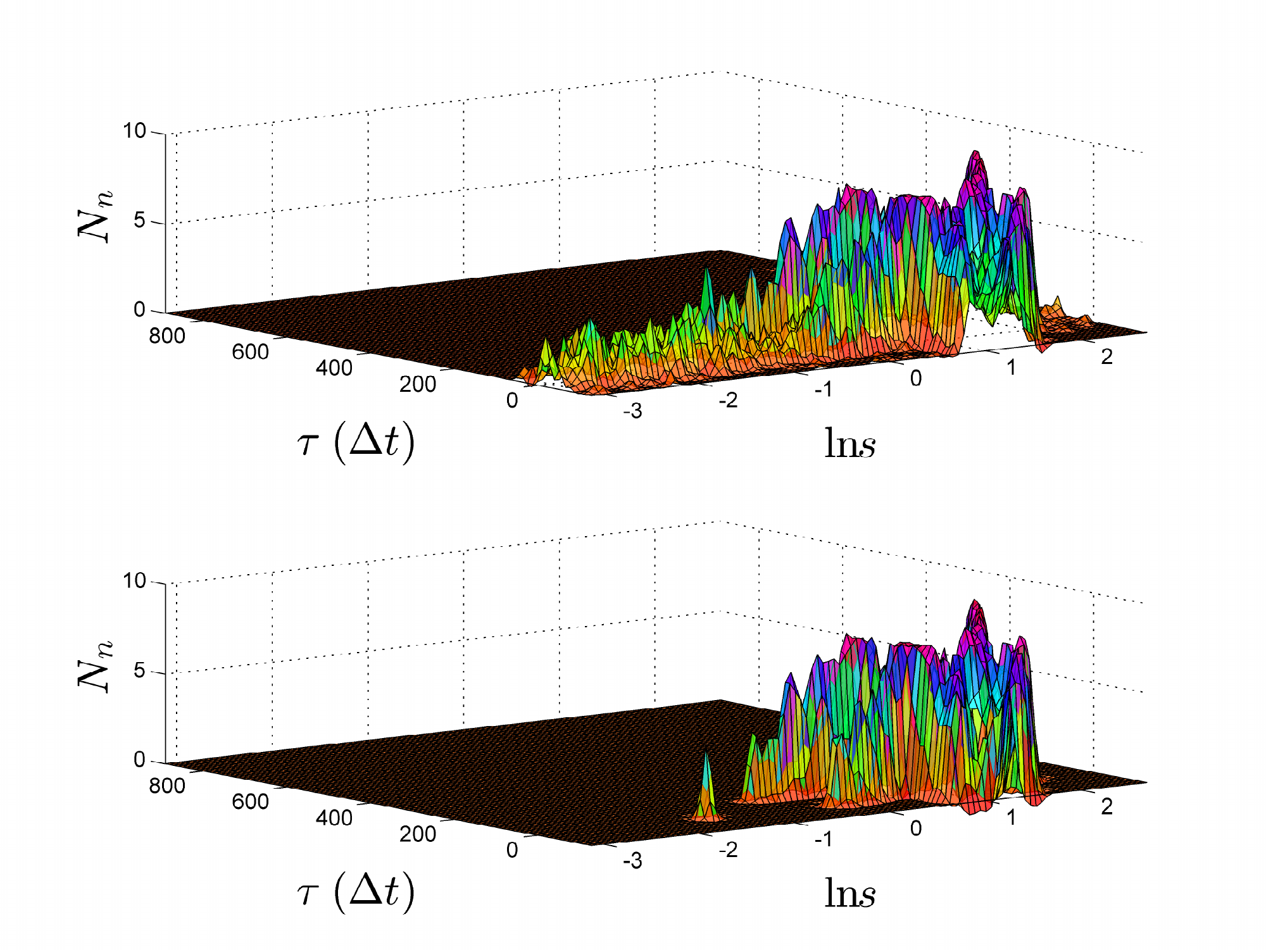}
\caption{\label{fig:ConnectivityMaps}The connectivity maps in the case of the original re-assigned wavelet transform (upper plot) and the final one (lower plot), obtained after discarding the information contained in the pixels 
with fewer than four important neighbours. $N_n$ denotes the number of important neighbours.}
\vspace{0.35cm}
\end{center}
\end{figure}

Applied to the $440$-Hz signal of the previous subsection, the linear correlation coefficient between the original and the reconstructed data increased from $0.998135$ (only the direct and the inverse wavelet transformations), 
to $0.998675$ (important contributions via the re-assigned wavelet transform ), to $0.999620$ (use of the connectivity plot). As seen in Fig.~\ref{fig:ConnectivityMaps}, the area of low connectivity, which is associated with 
high-frequency noise (small $s$ values), is removed after the application of the naive cut in the connectivity of each pixel. Although the appropriate processing of the re-assigned wavelet transform is postponed for the future, 
the involvement of the re-assignment method led to the reduction of the noise level present in the input signal, as did the (naive, in this work) processing of the re-assigned wavelet transform. These results are promising and 
a detailed study of the noise-related effects should be pursued.

\section{\label{sec:Conclusions}Conclusions}

The present paper provided the details of the first study in an ambitious research programme, aiming at the investigation of the employment of a wavelet-based analysis (WBA) in the processing of acoustical signals. If successful 
in this quest, the involvement of the wavelet transformations may serve as an alternative to the Fourier-based analysis (FBA) on which sound-processing commercial products mostly rely. The main advantage of the WBA (over the FBA) 
is the improvement in the signal quality at low and moderate frequencies.

The general expression for the Fourier transform of the wavelet, which is suitable for the processing of acoustical signals in the cochlea of the inner ear, was derived in Ref.~\cite{r1} on the basis of theoretical principles, 
as well as of the limitations in the localisation of a signal in the time and the frequency representations \cite{r2}; such wavelets were called `Reimann wavelets'. The Reimann mother wavelet was obtained in the time domain 
(Subsection \ref{sec:WaveletTimeDomain}) and was subsequently used in the processing involving the direct and the inverse wavelet transformations of Eqs.~(\ref{eq:EQ0011},\ref{eq:EQ0013}). In Subsection \ref{sec:SERe}, the 
structure equations, differential equations which the modulus and the phase of the wavelet transform fulfill, were introduced. Subsequently, the method of re-assignment, applied in Ref.~\cite{gm} to an analysis carried out within 
the framework of the short-time Fourier transformation, was adapted for an application in a WBA.

In Subsection \ref{sec:Parameters}, a scheme was set forth, enabling the determination of optimal values of the parameters of the Reimann wavelet on the basis of the goodness of the reproduction of a noise-free audio file 
containing a number of harmonic signals, ranging from $110$ to $3\,720$ Hz. Signals from a broader frequency domain were analysed in Subsection \ref{sec:Harmonic}; it was found that the wavelet transformations of 
Eqs.~(\ref{eq:EQ0011},\ref{eq:EQ0013}) successfully reproduce the input data in the frequency domain which is important in human acoustics.

Noise-related effects were discussed in Subsection \ref{sec:Noise}. Despite the fact that some noise is removed from the input data after applying the direct and the inverse transformations of Eqs.~(\ref{eq:EQ0011},\ref{eq:EQ0013}), 
the processing of the re-assigned wavelet transform is required in order to suppress the noise present in a signal. Although the preliminary results, obtained herein using a simple cut in the connectivity map, appear to be 
encouraging, a detailed study, addressing the subject of the noise suppression, is needed. To summarise in one sentence, the results obtained at this phase of the project are promising and further research on the application of 
the wavelet theory in the processing of acoustical signals should be pursued.

\begin{ack}
The author acknowledges helpful discussions with Hans-Martin Reimann and Peter Biller. Marcelo Magnasco has been helpful in clarifying a few questions on his implementation of the re-assignment method. Finally, the author is 
indebted to G{\"u}nther Rasche for his careful reading and valuable comments. This study has been financially supported by the `CTI - The Innovation Promotion Agency' of the `Federal Office for Professional Education and Technology 
(OPET)'; project title: `A new class of wavelets for an application to the auditory system', grant 11066.1. The paper is dedicated to Peter Biller (1962-2012).
\end{ack}

\newpage
\begin{table}[h!]
{\bf \caption{\label{tab:OptParVal}}}The optimal parameters of the Reimann wavelet, obtained after following the three-step procedure detailed in Subsection \ref{sec:Parameters}; the linear correlation coefficient $\rho$ between 
the input and the output values is also listed. The optimal parameter values are obtained on the basis of maximising $\rho$ and correspond to the output of the third step. The values of the settings for $N_M$, $d$, $\Delta s$, and 
$\Delta \tau$, used in this analysis, are given in the beginning of Subsection \ref{sec:Parameters}. The parameters $\nu$ and $c$ were kept fixed at $1$ throughout the optimisation.
\vspace{0.2cm}
\begin{center}
\begin{tabular}{|c|c|c|c|c|c|c|l|}
\hline
$\alpha / \pi$ & $\beta / \pi$ & $\phi_m / \pi$ & $\kappa$ & $\nu$ & $c$ & $\rho$ & Comment \\
\hline
$1.000$ & $8.500$ & $-2.000$ & $8.000$ & $1.000$ & $1.000$ & $0.999381$ & Initial input \\
$1.004$ & $8.607$ & $-1.682$ & $8.075$ & $1.000$ & $1.000$ & $0.999394$ & Output of the first step \\
$1.019$ & $8.726$ & $-1.694$ & $6.581$ & $1.000$ & $1.000$ & $0.999404$ & Output of the second step \\
$1.041$ & $8.851$ & $-1.831$ & $6.209$ & $1.000$ & $1.000$ & $0.999420$ & Output of the third step \\
\hline
\end{tabular}
\end{center}
\end{table}

\vspace{1cm}
\begin{table}[h!]
{\bf \caption{\label{tab:OrglAndRcstvsFrequency}}}The linear correlation coefficient $\rho$ between original and reconstructed data as a function of the frequency; harmonic signals have been used as input, spanning six octaves 
from $A_2$ to $A_8$ (Scientific pitch notation). The reconstruction is based on a mother wavelet obtained with the standard parameter values (see Table \ref{tab:OptParVal}).
\vspace{0.2cm}
\begin{center}
\begin{tabular}{|c|c|c|}
\hline
$f$ (Hz) & Note \cite{spn} & $\rho$ \\
\hline
$110$ & $A_2$ & $0.998678$ \\
$220$ & $A_3$ & $0.999195$ \\
$440$ & $A_4$ & $0.999860$ \\
$880$ & $A_5$ & $0.999822$ \\
$1\,760$ & $A_6$ & $0.999813$ \\
$3\,520$ & $A_7$ & $0.999836$ \\
$7\,040$ & $A_8$ & $0.995530$ \\
\hline
\end{tabular}
\end{center}
\end{table}

\end{document}